\documentstyle[preprint,aps]{revtex}	
\begin{document}
%
\preprint{DTP/93/52}
%
\title{Soft Gluon Radiation in Hadronic $t\bar t$ Production}
\author{V.~A.~Khoze and  J.~Ohnemus}
\address{
Department of Physics\\
University of Durham, Durham, DH1 3LE, England}
\author{W.~J.~Stirling}
\address{
Departments of Mathematical Sciences and Physics\\
University of Durham, Durham, DH1 3LE, England}
\maketitle
\begin{abstract}
We investigate soft gluon radiation in hadronic $t\bar t$ production. By
taking the top quark decay properly into account, we are able to study the
interplay of radiation both before and after the decay of the top quarks.
The production--decay and decay--decay radiative interferences depend
sensitively on the relative size of the gluon energy and the decay width.
Radiation patterns for various production mechanisms are compared.
\end{abstract}
\newpage
%
%
\narrowtext

\section{INTRODUCTION}

Discovery of the  top quark -- one of the basic components of the Standard
Model -- is one of the most important goals of present and future
high-energy colliders. At present, indirect evidence from electroweak
radiative corrections suggests a top quark mass between 100 and 200 GeV
(see for example reference \cite{LEPTOP}), and null results from searches
at the Fermilab $p \bar p $ collider yield a $95\%$~c.l. lower limit of
108~GeV and 103~GeV from the CDF \cite{CDFTOP} and D0 \cite{D0TOP}
collaborations, respectively. Optimistically, the top quark will be
discovered at the Fermilab  $p \bar p $ collider in the next few years.

The two most important properties of the top quark, which can be measured
once its existence is established, are the mass $m_t$ and the decay width
$\Gamma_t$. The first of these is a fundamental parameter of the theory,
while the second provides  an opportunity to check the existence of any
non-Standard Model  decay channels. However, obtaining  precise
measurements of these parameters at a hadron collider is a far from easy
task. In particular, a detailed knowledge of the structure of the final
state is necessary. For example, gluon radiation from the initial state,
from the produced top quarks, and from the top quark decay products  can
influence the reconstruction of the top mass from its decay products. Soft
gluon radiation, in particular the interference between radiation before
and after top decay, also depends sensitively  on the decay width
\cite{KOS,JIKIA}.

In this paper we study the \lq antenna pattern' of  soft gluon radiation in
$t \bar t$ production in high energy hadron-hadron collisions.  In
particular, we are interested in (a) the process dependence ($q \bar q \to
t \bar t$ {\it vs.} $ gg \to t \bar t$ ) of this radiation and (b) the
effect of the top quark instability on the radiation pattern. The former
depends on the colour topology of the different processes, while the latter
induces dependence on $\Gamma_t$ and on the orientation of the top decay
products. Our study builds on earlier work. In reference \cite{KOS}, for
example, the colour-singlet production process $e^+e^-\to t\bar t \to b W^+
\bar b W^-$ was analyzed in detail, and a gauge-invariant approach for
identifying  and calculating the various production and decay contributions
to the radiation pattern  was presented. It is the extension of these ideas
to the more complicated processes $q \bar q , gg\to t\bar t \to b W^+ \bar
b W^-$ which constitutes the main goal of the present study. The radiation
pattern for {\it stable} top production $q \bar q , gg\to t\bar t$ has been
analyzed in detail  in reference \cite{KELLIS,MW} (see also \cite{BOOK}).
Again, our purpose here is to extend this analysis to the experimentally
more relevant $b W^+ \bar b W^-$ final state.
Note that in our analysis the soft gluons are sufficiently energetic
so that a
perturbative treatment is valid.
Non-perturbative effects, including possible long distance
interactions
with the spectator partons, have been discussed in Ref.~\cite{ORR}.

Our aim, then, is to give a complete and systematic account of soft gluon
radiation in hadronic $t \bar t$ production, accounting for colour topology
of the different processes, decay product kinematics, and finite top width
effects. In the next section, we describe a method for writing down the
antenna pattern for {\it any} $\;t \bar t$  production process of the form
$A B \to t\bar t \to b W^+ \bar b W^-$. Although our main interest here is
in the strong interaction processes $q \bar q, gg \to t \bar t$, we also
present results for $q \bar q \to Z^* \to t \bar t$ and $gg \to H \to t
\bar t$. These processes, with their unique colour structure, are useful
for comparison. In Section III we present some numerical results. In
particular, we consider final states with particular (and experimentally
typical) orientations of the final state $b$ quarks and $W$ bosons. We
study the soft gluon antenna patterns, focusing on the process and top
width dependence. Our objective is not to emulate a Monte Carlo
analysis including realistic detector acceptances, but rather to emphasize
the most important features of the soft gluon radiation accompanying $ t
\bar t $ production and decay  which should be taken into account in more
realistic studies.   Finally, in Section IV we summarize our   main
results.

\section{FORMALISM}

In the soft gluon limit, the radiation pattern for {\it any} elementary
$2\to 2$ hard scattering process $f$ can be written as \cite{BOOK}
\begin{eqnarray}
\overline{\sum} |{\cal M}^{(f)} |^2 =
\sum_{ij} \; \widehat{ij} \; A^{(f)}_{ij} \; ,
\end{eqnarray}
where the sum $i,j$ is over the external particles.
The `antenna' $\widehat{ij}$ is defined by
\begin{eqnarray}
\widehat{ij} \; \equiv  \; {p_i\cdot p_j \over p_i\cdot k \ p_j\cdot k} \; ,
\end{eqnarray}
where $p_i$ and $p_j$ are the four-momentum vectors of particles $i$ and
$j$ and $k$ is the gluon four-momentum vector. The coefficients
$A^{(f)}_{ij}$ are functions of the kinematic invariants of the $2\to 2$
process. The simplest example is the process $e^+e^- \to q \bar q (g)$
where, for massless quarks, there is only one such antenna and
$\overline{\sum} |{\cal M}|^2 \propto \widehat{q \bar q}$.

In the present context we are interested in the following $t\bar t$
production processes:
\begin{mathletters}
\begin{eqnarray}
&&e^- e^+          \to t \bar t \>, \\
&&q \bar q \to Z^* \to t \bar t \>, \\
&&gg       \to H   \to t \bar t \>, \\
&&q \bar q         \to t \bar t \>, \\
&&gg               \to t \bar t \>.
\label{EQ:TT}
\end{eqnarray}
\end{mathletters}
(Note that the soft gluon radiation pattern for the  process $\gamma \gamma
\to t \bar t$ is identical to the pattern for the $e^- e^+ \to t \bar t$
process.) Labeling the initial and final state momenta by
\begin{eqnarray}
A(k_1) + B(k_2) \to  t(q_1) + \bar{t}(q_2) \to
b(p_1) + W^+ + \bar{b}(p_2) + W^- \>,
\end{eqnarray}
the general form of the differential distribution for soft
gluon radiation $(k^\mu)$ is \cite{KOS}
\begin{eqnarray}
{1\over d\sigma_0}\ {d\sigma\over d\omega_g\> d\cos\theta_g \> d\phi_g}\ = \
{\alpha_s\over 4 \pi^2} \ \omega_g \ {\cal F} \; ,
\label{EQ:dN}
\end{eqnarray}
where $d\sigma_0$ is the differential cross section for the lowest order
process ({\it i.e.}, with no gluon radiation), $\omega_g$ is the energy of
the soft gluon, and $\alpha_s$ is the strong  coupling. For the case of
{\it stable} top quarks and massless initial state particles, the
distribution function ${\cal F}$ has the general form
\begin{eqnarray}
{\cal F}_0  = c_1 \widehat{k_1 k_2} + c_2 \widehat{k_1 q_1}
+ c_3 \widehat{k_1 q_2} + c_4 \widehat{k_2 q_1}
+ c_5 \widehat{k_2 q_2} + c_6 \widehat{q_1 q_2} + c_7 \widehat{q_1 q_1}
+ c_8 \widehat{q_2 q_2} \>,
\label{EQ:GENERAL}
\end{eqnarray}
where the antennae are defined as before by
\begin{eqnarray}
\widehat{pq} \> \equiv \> {p\cdot q \over p\cdot k \ q\cdot k} \; .
\end{eqnarray}
The coefficients $c_i$ are given in Table~\ref{TAB:COLOUR} for the five
$t\bar t$ production processes listed in Eq.~(3). The functions $X$ and $Y$
appearing in the $gg\to t \bar t$ process  are discussed later.

The general method for treating {\it unstable} heavy particles such as  the
top quark has been presented in Ref.~\cite{KOS}.  The top quark {\it decay}
is included by letting  ${\cal F}_0 \to {\cal F}$, with ${\cal F}$ given by
Eq.~(\ref{EQ:GENERAL}) with the replacements:
\begin{eqnarray}
\widehat{k_1 k_2}   &\to& \widehat{k_1 k_2}  \>, \nonumber \\
\widehat{k_1 q_1}   &\to& \widehat{k_1 q_1}
+ \chi_1\; [ \widehat{k_1 p_1} -\widehat{k_1 q_1} ]   \>, \nonumber \\
\widehat{k_1 q_2}   &\to& \widehat{k_1 q_2}
+ \chi_2\; [ \widehat{k_1 p_2} -\widehat{k_1 q_2} ]   \>, \nonumber \\
\widehat{k_2 q_1}   &\to& \widehat{k_2 q_1}
+ \chi_1\; [ \widehat{k_2 p_1} -\widehat{k_2 q_1} ]   \>, \nonumber \\
\widehat{k_2 q_2}   &\to& \widehat{k_2 q_2}
+ \chi_2\; [ \widehat{k_2 p_2} -\widehat{k_2 q_2} ]  \>,
\label{EQ:REPS} \\
\widehat{q_1 q_2}   &\to& \widehat{q_1 q_2}
+ \chi_2\; [ \widehat{q_1 p_2} -\widehat{q_1 q_2} ]
+  \chi_1 \; [ \widehat{q_2 p_1} -\widehat{q_1 q_2} ]   \nonumber \\
& & \ + \ \chi_{12} \; [ \widehat{p_1 p_2} - \widehat{q_1 p_2}
- \widehat{q_2 p_1} + \widehat{q_1 q_2} ]   \>,
\nonumber \\
\widehat{q_1 q_1} & \to & 2 \widehat{q_1 q_1} + \widehat{p_1 p_1}
-2\widehat{q_1 p_1} + 2 \chi_1 \; [ \widehat{q_1 p_1}
- \widehat{q_1 q_1} ]   \>, \nonumber \\
\widehat{q_2 q_2} &\to& 2 \widehat{q_2 q_2} + \widehat{p_2 p_2}
- 2\widehat{q_2 p_2}
+ 2 \chi_2\;  [ \widehat{q_2 p_2} -\widehat{q_2 q_2} ] \> .
\nonumber
\end{eqnarray}
We make the following comments concerning the above results.

\begin{itemize}

\item[{(i)}] For unstable top quarks, we see that there are additional
contributions from antennae involving the daughter $b$~quarks,
corresponding to gluon emission both before {\it and} after the $t$~quarks
decay. The profile functions $\chi$, which depend on the top width
$\Gamma_t$ and determine the size of the interference effects, are given by
\cite{KOS}
\begin{eqnarray}
\chi_i & = & {m_t^2\Gamma_t^2 \over (q_i\cdot k)^2 + m_t^2 \Gamma_t^2 } \>,
\label{EQ:CHII} \\
\noalign{\vskip 5pt}
\chi_{12} & = & {m_t^2\Gamma_t^2\; (q_1\cdot k\; q_2\cdot k + m_t^2\Gamma_t^2)
 \over \Bigl[ (q_1\cdot k)^2 + m_t^2 \Gamma_t^2 \Bigr]
    \; \Bigl[ (q_2\cdot k)^2 + m_t^2 \Gamma_t^2 \Bigr] } \; .
\label{EQ:CHI}
\end{eqnarray}
These functions have the (formal) property that $\chi_i, \chi_{12} \to 0$
as $\Gamma_t \to 0$, and $\chi_i, \chi_{12} \to 1$
as $\Gamma_t \to \infty$. In the former limit, we have
\begin{eqnarray}
{\cal F} \to {\cal F}_0 + c_7 \bigl[ \widehat{q_1 q_1}
+ \widehat{p_1 p_1} -2\widehat{q_1 p_1} \bigr]
+ c_8 \bigr[ \widehat{q_2 q_2}+ \widehat{p_2 p_2}
- 2\widehat{q_2 p_2}  \bigr] \; ,
\end{eqnarray}
which is the stable top result together with additional $\widehat{t\, b}$
and  $\widehat{\bar{t}\, \bar{b}}$ antennae \cite{KOS}. In the other
extreme, $\chi_i, \chi_{12} \to 1$, the top quarks decay immediately after
production and  the radiation pattern is identical to that for the process
\begin{eqnarray}
A(k_1) + B(k_2) \to  b(p_1) + \bar{b}(p_2)
\end{eqnarray}
at reduced invariant squared energy $W^2 = (p_1 + p_2 )^2$ in the final
state.

\item[{(ii)}] The functions $X$ and $Y$ appearing in the coefficients
for the $gg \to t \bar t$ process are given by
\begin{eqnarray}
X & = & {N^2\over 4C_F}\; \left[ (1+2\mu)\left( {1\over U} - {1\over T}\right)
- \mu^2  \left( {1\over U^2} - {1\over T^2}\right) + 2(U-T) \right] \nonumber
\\
& & \times \; \left[ {1\over UT} - {N\over C_F}  \right]^{-1}
 \;  \left[ T^2+U^2  + 2 \mu -{\mu^2\over UT} \right]^{-1}   \; ,
\end{eqnarray}
and
\begin{eqnarray}
Y & = & {1\over 4C_F}\; \left[{1\over N^2UT} + 2   \right]
 \; \left[ {1\over UT} - {N\over C_F}  \right]^{-1}\; ,
\end{eqnarray}
where
\begin{eqnarray}
T= {k_1\cdot q_1 \over k_1\cdot k_2}, \qquad
U= {k_1\cdot q_2 \over k_1\cdot k_2}, \qquad
\mu= {m_t^2 \over   k_1\cdot k_2} \; .
\end{eqnarray}
(The $X$ and $Y$ functions here are closely related to those given in Ellis
and Sexton \cite{KELLIS}; here we have normalized $X$ and $Y$ by dividing
them by the $gg \to t \bar t$ squared amplitude.) Figure~\ref{FIG:XYFUN}
shows the functions $X$ and $Y$ plotted versus the  $t$ quark scattering
angle (in the $gg$ centre-of-mass frame) for three different beam energies.
Note that at threshold $T=U=\mu=1/2$ and
\begin{eqnarray}
X = 0,\qquad  Y = {1\over 2N^2} \; {2+N^2\over 4C_F - N} =
{11 \over 42} \quad (N=3) \>.
\end{eqnarray}
The numerically small values of $Y$ [$Y/X \sim {\cal O}(1/N^2)$],  together
with the fact that collinear singularities are absent from the
corresponding combination of antennae, leads to  a negligible contribution
from this piece in practice \cite{KELLIS,MW}. The contributions from the
$t\bar t$ antenna are not suppressed by $1/N^2$ in processes 3a, b, and c,
where the $t\bar t$ pair is produced in a colour singlet state.  In
processes 3d and e, the azimuthal asymmetry of emission off $t$-quarks is
violated even in the large $N$ limit, because of the presence of the
$\widehat{k_i q_j}$ antennae.

\item[{(iii)}] Although the bulk of the coefficients in
Table~\ref{TAB:COLOUR} are simply colour factors, independent of the
subprocess invariants, the cross section for soft gluon radiation does {\it
not} in general factorize into an eikonal factor times the lowest-order
cross section.  This lack of factorization is the origin of the $X$ and $Y$
terms appearing in Table~\ref{TAB:COLOUR} for the $gg \to t \bar t$
process. The reason can be traced to the colour structure of the $2\to 2 $
process \cite{MW,BOOK}.  For all but the $gg\to t \bar t$ process, the
colour flow at the amplitude level can be uniquely represented by
continuous lines (\lq strings') linking the external particles. Thus
Figs.~\ref{FIG:COLOR}(a) and (b) show the colour flow for $e^-e^+\to t \bar
t$ and $q \bar q \to t \bar t$, respectively. Note that  in the leading
$1/N$ limit (the planar limit) there is a one-to-one correspondence between
the colour flow lines and the non-zero antenna coefficients. In contrast,
the $gg\to t \bar t$ process  has two possible colour flows at the
amplitude level as  shown in Figs.~\ref{FIG:COLOR}(c) and (d).  It is the
interference between these two contributions that spoils the identification
of the colour flows in the planar limit. However, this interference is
suppressed in the large $N$ limit, where  we find
\begin{eqnarray}
 {\cal F}  &\to&
{1\over 4}\; (N-2X)\; \left[ 2  \widehat{k_1 q_1} + 2 \widehat{k_2 q_2}
+ 2\widehat{k_1 k_2}
-  \widehat{q_1 q_1} - \widehat{q_2 q_2}   \right] \nonumber \\
&+&
{1\over 4}\; (N+2X)\; \left[  2 \widehat{k_1 q_2} + 2 \widehat{k_2 q_1}
+ 2 \widehat{k_1 k_2}
-  \widehat{q_1 q_1} - \widehat{q_2 q_2}   \right] \; ,
\end{eqnarray}
which is the separation into two positive-definite colour structures
discussed in Ref.~\cite{MW}.

\item[{(iv)}]In the large $N$ limit, the  $\widehat{tb}$ and  $\widehat{\bar t
\bar b}$ antennae can interfere only in the case of the first three
processes in Eq.~(3).  For processes 3d and 3e, in this limit, the decays are
uncorrelated in colour and the interference is absent.
\end{itemize}

\section{NUMERICAL RESULTS}

In the previous section we have seen how the different colour structure of
the various processes leads to different weightings  of the antennae
contributions. In this section we study the  implications of this for the
actual gluon distribution.  Our purpose is not to give a comprehensive
treatment  of all the possibilities, but rather to consider some simple
configurations which illustrate the underlying physics.

In studying the way in which the various antennae contribute, it is
instructive to consider all five processes listed in Eq.~(3).   Since we
may neglect the soft gluon  in the kinematics,  the $t$- and $\bar
t$-momenta  are back-to-back in the centre-of-mass frame of the collision.
The  $b$- and $\bar b$-momenta are also taken to be back-to-back.  This is
for simplicity only --   the qualitative features are unchanged as long as
the $b \bar b$ angle   is not chosen particularly small.

Distributions of the soft gluon radiation in the polar and azimuthal angles
(relative to the beam direction) are presented  for a soft gluon with
energy $\omega_g = 5$~GeV. Again, this value is not  critical: what matters
as far as the interference contributions are concerned is the relative size
of $\omega_g$ and $\Gamma_t$.  Our results should be interpreted as
referring to the distribution of a soft hadronic jet accompanying the
significantly more energetic  $b W^+ \bar b W^-$ particles.  The
distributions are calculated at the parton level and are plotted in the
parton subprocess  centre-of-mass frame. Unless otherwise stated, we assume
a parton beam energy of $E_{\rm beam} = 2 m_t$, motivated by the fact that at
hadron colliders the average  subprocess centre-of-mass energy is several
times the threshold energy for $t \bar t$ production. A realistic study of
soft gluon radiation in hadronic $t\bar t$ production would of course
require a convolution of the parton-level cross sections with parton
distribution functions.  Such a convolution would smear the results
illustrated here, although one could  in principle reconstruct this frame
directly  if all the top quark decay products were measured.

The following short-hand notation will be used to denote the polar
($\theta_g$) and azimuthal angle ($\phi_g$) distributions:
\begin{eqnarray}
{dN \over d\cos \theta_g} &\equiv& \omega_g \, {1\over d\sigma_0} \,
{d\sigma \over d\omega_g d\cos \theta_g d\phi_g}
\Biggl|_{\phi_g = 0^{\circ} \phantom{{\rm GeV}} \atop \omega_g = 5\; {\rm GeV}}
= {\alpha_s \over 4 \pi^2} \, \omega_g^2 \, {\cal F}(\theta_g) \>, \\
\noalign{ \hbox{\rm and} }
{dN \over d\phi_g} &\equiv& \omega_g \, {1\over d\sigma_0} \,
{d\sigma \over d\omega_g d\cos \theta_g d\phi_g}
\biggl|_{\theta_g = 90^{\circ} \phantom{{\rm GeV}}
\atop \omega_g = 5\;{\rm GeV}}
= {\alpha_s \over 4 \pi^2} \, \omega_g^2 \, {\cal F}(\phi_g) \>.
\end{eqnarray}
Recall that in the soft gluon approximation, the gluon energy distribution
falls as $1/\omega_g$. Thus an extra factor of $\omega_g$ has been included
in the above definitions to make the distributions dimensionless
quantities.   The numerical results presented here have been generated
using the following input parameters:  $m_t = 140$~GeV, $m_b = 5$~GeV,
$M_W^{} = 80$~GeV, and $\alpha_s = 0.1$. For all the distributions, the
coordinate system is chosen such that the three-momenta of the incoming
partons are along the $\pm \hat z$ directions. Furthermore, the
three-momenta of the outgoing $t$ and $\bar t$ are chosen to be in the
$+\hat x$ and $-\hat x$ directions, respectively. For the
$dN/d\cos\theta_g$ distributions, the three-momenta of the $b$ and $\bar b$
quarks are chosen to be back-to-back in the $xz$-plane, with the $b$-quark
three-momentum vector $45^{\circ}$  away from the $t$-quark three-momentum
vector as illustrated in Fig.~\ref{FIG:XYZPLANE}(a). The soft gluon
three-momentum is also taken to be in the $xz$-plane, and thus the
three-momenta of all the partons lie in the $xz$-plane as depicted in
Fig.~\ref{FIG:XYZPLANE}(a).   For the $dN/d\phi_g$ distributions, the
three-momenta of all the final state partons are taken to be in the
$xy$-plane, which is transverse to the momenta of the incoming partons; see
Fig.~\ref{FIG:XYZPLANE}(b).

We begin by presenting results for the polar angle distribution of the soft
gluon.  Figure~\ref{FIG:A}  shows $dN/d\cos\theta_g$ versus $\theta_g$ for
incoming partons with energy $E_{\rm beam} = 2m_t = 280$~GeV,  where $m_t =
140$~GeV is the top quark mass.  The Standard Model Born value $\Gamma_t =
\Gamma_{\rm sm}^{(0)} = 0.67$~GeV for $m_t=140$~GeV,  has been used for the
top quark width. The soft gluon radiation patterns for the five processes
listed in  Eq.~(3) are shown in the figure.  The angles of the incoming and
outgoing  partons are indicated at the top of the figure.  The divergence
in the radiation pattern at $\theta_g = 0^\circ$ and $180^\circ$ (and
$360^\circ$)  for the hadronic production processes is the initial state
collinear singularity, {\it i.e.}, $ q \to q+g$, $g \to g+g$. The strength
of the singularity  simply reflects the colour charge of the incoming
particles.  For example, the radiation pattern in the region $\theta_g
\approx 0^\circ$  is dominated by the antennae containing a factor of
$k_1$. If $k$ is almost parallel to $k_1$ then
\begin{eqnarray}
\widehat{k_1 a} \approx {2\over \omega_g^2}\;{1\over \theta_g^2} \>,
\label{EQ:COL}
\end{eqnarray}
independent of the other momentum ($a^\mu$)  in the antenna. There are
three such terms on the right-hand-side of Eq.~(\ref{EQ:GENERAL}), and
hence
\begin{eqnarray}
{\cal F} \approx {2\over \omega_g^2}\;
{c_1+c_2+c_3\over \theta_g^2} \>.
\end{eqnarray}
 From Table~\ref{TAB:COLOUR}, we see that $c_1+c_2+c_3  = 2C_F\ (2N)$ for
the $q \bar q\ (gg)$ initiated processes, respectively. Thus near $\theta_g
= 0^\circ$ and $180^\circ$ the curves for the $gg$ initiated processes are
larger than those for the $q \bar q$ initiated processes, while the
non-singular  $e^-e^+$ distribution is very small.

In the same way, all of the processes exhibit  a sharp dip in the radiation
pattern in the direction of the $b$ and $\bar b$ quarks when $m_b/E_b \ll 1$.
These dips are the well known  heavy-quark dead cones \cite{DEAD}.
For example, close to the $b$-quark direction we have [{\it cf.},
Eq.~(\ref{EQ:COL})]
\begin{eqnarray}
\widehat{p_1 a} \approx {2\over \omega_g^2}\;{\tilde\theta_g^2\over
\bigl[ \tilde\theta_g^2 + (m_b/E_b)^2 \bigr]^2} \; ,
\label{EQ:COLB}
\end{eqnarray}
where now $\tilde\theta_g$ is the angle between the gluon and  $b$-quark
directions. The angular width of these dead cones is proportional to
$m_b/E_b$. Again, it is straightforward to check that the dead-cone
behavior near the  $b$-quark directions is the same for all processes. This
follows from the fact that $c_2+c_4+c_6  =-2 c_7  = 2C_F$  (see
Table~\ref{TAB:COLOUR}).

The corresponding dead cones about the $t$ and $\bar t$ quark directions
are much broader due to the larger top quark mass.  The minimum of the $t
\> (\bar t)$ quark dead cone does not coincide with the $t \> (\bar t)$
quark direction, instead, the minimum is shifted to a slightly larger
angle.   This is because the $\widehat{q_1 p_1}$ and $\widehat{q_2 p_2}$
terms in Eq.~(\ref{EQ:REPS}) give a contribution which is a decreasing
function of $\theta_g$ in the neighborhood around the $t \> (\bar t)$ quark
direction, whereas the contribution from the other terms has a minimum in
the $t \> (\bar t)$  quark direction.   The resulting sum thus has a
minimum at a value of $\theta_g$  that is slightly larger than the
direction of the $t \> (\bar t)$ quark. The other interesting feature to
note  in Figure~\ref{FIG:A} is the difference between the distributions for
the $q \bar q \to Z^* \to t \bar t $ and  $q \bar q \to t \bar t $
processes. The presence of the  $\widehat{k_1 q_1}$ and $\widehat{k_2 q_2}$
antenna for the latter {\it increases} the radiation between the $q$ and
$b$ quarks, and {\it decreases} that between the $b$ and $\bar q$ quarks,
relative to the former process. The effect is clearly seen in the dips on
either side  of the $b$-quark dead cone in Figure~\ref{FIG:A}.

Figure~\ref{FIG:C} shows $dN/d\cos\theta_g$ versus $\theta_g$  for incoming
partons with energy $E_{\rm beam} = 2m_t = 280$~GeV as in
Fig.~~\ref{FIG:A}, but now with the top quark width taken to the formal
limit $\Gamma_t \to \infty$.   In this limit, the interference factors
$\chi_1, \chi_2$, and $\chi_{12}$ in Eqs.~(\ref{EQ:CHII}) and
(\ref{EQ:CHI}) all go to 1.  This limit corresponds to the top quark
decaying instantly, thus the resulting radiation pattern is equivalent to
the radiation pattern for $b \bar b$ production.  In particular,  the
radiation pattern is  independent of the $t \> (\bar t)$  direction.
Comparing the case of $\Gamma_t = \Gamma_{\rm sm}^{(0)}$ in
Fig.~\ref{FIG:A} with the present case of $\Gamma_t = \infty$, we see that
with the exception of the $e^-e^+$ process, the  radiation is again not
symmetric around the $b \> (\bar b)$ dead cone. However, the maximum now
appears on the side of the cone  closest to the incoming beam direction.
For the reasons noted above, the asymmetry is particularly strong for the
$q \bar q \to t \bar t$ process.   We also see that in the regions where
the $t \> (\bar t)$ dead cones were  previously located ($\theta_g \approx
100^\circ$ and $280^\circ$), the $e^- e^+ \to t\bar t$ and $q\bar q \to t
\bar t$ curves have changed considerably, whereas the $gg \to H \to t\bar
t$ and $gg \to t \bar t$ curves have changed little.  For these latter
processes, the radiation pattern receives a larger contribution from
radiation off the initial-state $gg$, which  is of course independent of
$\Gamma_t$. On the other hand, the radiation patterns for the $e^- e^+ \to
t \bar t$ and $q\bar q \to t \bar t$ processes are  more sensitive to the
radiation from the final state quarks,  which is very different for the two
cases of $b\bar b$ production versus $t \bar t$ production and decay.

The variation of the radiation pattern with the top quark width is
illustrated in Fig.~\ref{FIG:E} for the process $gg \to t \bar t$.  As
before,  the distribution $dN/d\cos\theta_g$ is plotted versus $\theta_g$
for incoming partons with energy $E_{\rm beam} = 2m_t = 280$~GeV.  Curves
are shown for four values of the top quark width: $\Gamma_t = \infty$,
5~GeV, 0.67~GeV (the Standard Model Born value), and 0.  The radiation
pattern for the Standard Model Born top quark  width is very close to the
radiation pattern for $\Gamma_t = 0$. Between the incoming $g(k_1)$ and the
$b$-quark, the $\widehat{k_1 p_1}$ antenna causes the radiation to increase
as $\Gamma_t$ (and hence $\chi_1$) increases. On the other side of the
$b$-quark dead-cone, the effect of emission from the top quark is seen to
decrease as $\Gamma_t$ increases. On the whole, however, the curves show
that the $gg \to t\bar t$ process is not very sensitive to the top quark
width. The $q\bar q \to t\bar t$ process on the other hand is much more
sensitive to the top quark width (see Figs.~\ref{FIG:A} and \ref{FIG:C}).

A more graphic representation of the effects described above can be
obtained by considering the radiation pattern in polar coordinates. Thus
Fig.~\ref{FIG:G} shows $dN/d\cos\theta_g$  versus $\theta_g$, with the
distance from the origin proportional  to the magnitude of the
distribution. As before, all particles lie in the  $xz$-plane; their
directions are indicated by arrows on the figure.    The radiation pattern
for the process $gg \to t\bar t$ is shown for  two values of the top quark
width: the Standard Model Born value and the $\Gamma_t \to \infty$ limit
which  is equivalent to $b \bar b$ production. (Note that this is simply a
different representation of the same two curves in the previous  figure.)
The collinear singularities along the $\pm \hat z$ directions are clearly
evident, as are the dead cones  about the $b \> (\bar b)$ and $t \> (\bar
t)$ quark directions. Notice again that the asymmetry of the lobes about
the $b$-quark direction  is different for the two cases, for the reasons
discussed earlier.

The polar angle distributions presented above emphasize the contributions
involving radiation from the incoming particles, {\it i.e.}, the antennae
involving $k_1$ and $k_2$. To focus on the radiation from the final state
particles it is useful to consider the azimuthal angle distribution of the
soft gluon radiation.  The three-momenta of the final state partons are now
fixed  to lie in the plane transverse to the incoming partons.   In this
configuration, the effects of the initial state radiation are  minimized.
We might therefore  expect the $dN/d\phi_g$ distributions to be more
similar than the $dN/d\cos \theta_g$ distributions for the five processes
listed in  Eq.~(3).  Figure~\ref{FIG:H}  shows the azimuthal distribution
of soft gluon radiation when all the outgoing partons  are in the
transverse plane.  Distributions are again shown for the five processes
listed in Eq.~(3). The directions of the three-momenta of the outgoing
heavy quarks are indicated at the top of the figure. The momenta of the $t$
and $\bar t$ quarks are at $0^\circ$ and $180^\circ$, respectively, while
the $b$ and $\bar b$ momenta are at $90^\circ$  and $270^\circ$,
respectively.   The dead cones in the directions of the $t$ and $b$ ($\bar
t$ and $\bar b$) quarks are clearly evident in this figure.   The width of
the dead cones is proportional to $m_Q/E_Q$, and thus the  $t$ ($\bar t$)
quark dead cone is much wider than the $b \> (\bar b)$ quark dead cone.
The minimum in the $t \> (\bar t)$ dead cone  does not coincide with the $t
\> (\bar t)$ direction, instead, the minimum is shifted to a slightly
smaller angle.  The reason for this shift is the same as discussed for the
$dN/d\cos \theta_g$ distribution in Fig.~\ref{FIG:A},  except that now the
$\widehat{q_1 p_1}$ and $\widehat{q_2 p_2}$ terms in Eq.~(\ref{EQ:REPS})
give a contribution which is an increasing function of $\phi_g$ in the
neighborhood around the direction of the $t \> (\bar t)$ quark. The
contribution from the other terms has a minimum in the  $t \> (\bar t)$
direction and is symmetric about this direction. The resulting sum now has
a minimum at a value of $\phi_g$ that is slightly smaller than the $t \>
(\bar t)$ quark direction. The $\widehat{tb}$ antennae, {\it i.e.}, the
$\widehat{q_1 p_1}$ and $\widehat{q_2 p_2}$ terms in Eq.~(\ref{EQ:REPS}),
are also responsible for the lack of symmetry in the radiation pattern
about the $t$ and $b$ quark directions. Note that the ordering of the
different processes is preserved over the  complete $\phi_g$ range, with
the effects of the initial radiation providing an approximately constant
background  to the dominant contributions  involving the $t$ and $b$
quarks. The ordering reflects the colour charges of the initial particles,
$gg > q \bar q > e^-e^+$. Note, however, that the hadronic production
processes are suppressed relative to the  corresponding $s-$channel colour
singlet processes,  {\it e.g.}, $(q \bar q \to Z^* \to t \bar t) \; > \: (q
\bar q \to t\bar t)$. The reason, evident in Table~\ref{TAB:COLOUR}, is the
suppression of the $\widehat{t \bar t}$ antenna (the coefficient $c_6$) for
the hadronic production processes.

Figure~\ref{FIG:I} is the same as Fig.~\ref{FIG:H},  except that now the
$b$ and $\bar b$ momenta have been fixed at $45^\circ$ and $235^\circ$,
respectively. The $b \> (\bar b)$ quark is now closer to the $t \> (\bar
t)$ direction, consequently, the $b \> (\bar b)$ quark is now moving
faster, and thus the  $b \> (\bar b)$ quark dead cone is now narrower than
in the previous figure. If the trend of moving the $b \> (\bar b)$ quark
closer to the  $t \> (\bar t)$ quark direction continues, the $b \> (\bar
b)$ quark continues to move faster and its dead cone continues to narrow.
In the limiting case of the $t$ and $b$ ($\bar t$ and $\bar b$) quarks
parallel, the dead cone is very narrow and the radiation pattern is
symmetric about  the $t\, b$ and $\bar t \, \bar b$ directions.

Figure~\ref{FIG:L} is the same as Fig.~\ref{FIG:H},  but with the top quark
width set to infinity.  This corresponds to the top quark decaying
instantly, and  so the radiation pattern is equivalent to that of $b\bar b$
production.  The major difference between this figure and Fig.~\ref{FIG:H}
is the lack of the top quark dead cones in the present figure. The
radiation pattern is now symmetric about the $b$ and $\bar b$ directions.
(The $t$ and $\bar t$ directions are irrelevant  since the $t$ and $\bar t$
decay instantly).  Notice that the magnitude of the $q \bar q \to t \bar t$
curve is now less than  the $e^- e^+ \to t \bar t$ curve. The reason for
this is that in the  limit $\Gamma_t \to \infty$ ($\chi_{12} \to 1$) the
$\widehat{p_1 p_2}$ antenna (negative) contribution is enhanced.
The colour suppression
of this contribution in the $q \bar q \to t \bar t$ case is no longer
compensated by the  background initial state radiation.

The variation of the azimuthal radiation pattern with the top quark width
is illustrated in Fig.~\ref{FIG:N} for the process $gg\to t\bar t$. The
distribution $dN/d\phi_g$ is plotted versus $\phi_g$ for incoming partons
with energy $E_{\rm beam} = 2m_t = 280$~GeV.  The radiation pattern is
shown for four values of the top quark width: $\Gamma_t = \infty$, 5~GeV,
0.67~GeV (the Standard Model Born value), and 0.  The radiation pattern for
the Standard Model Born top quark width is almost identical to the
$\Gamma_t = 0$ pattern.  The dead cone about the $t \> (\bar t)$ quark is
obscured already for $\Gamma_t = 5$~GeV.

In Fig.~\ref{FIG:P} we plot $dN/d\phi_g$ versus $\phi_g$  in polar
coordinates. The radiation pattern for the process $gg \to t\bar t$ is
shown for  $\Gamma_t = \Gamma_{\rm sm}^{(0)}$ (solid line)  and $\Gamma_t =
\infty$ (dashed line). The angle between the momenta vectors of the $t$ and
$b$ quarks is $0^\circ, 45^\circ, 90^\circ, 135^\circ$, and $180^\circ$,
for parts (a), (b), (c), (d), and (e), respectively.   The radiation
patterns are quite different for the two cases.  The $b\bar b$ radiation
pattern (dashed line) is always symmetric about the $b \bar b$ direction.
(The $t\bar t$ direction is irrelevant in this case.) For the Standard
Model width (solid line), the radiation is always largest in the angular
sector between the $t$ and $b$ ($\bar t$ and $\bar b$) directions.  For
example, in part (c), the radiation in the first and third quadrants is
much larger than in the second or fourth quadrants.  This is a consequence
of the colour antenna connecting the $t$ and $b$ ($\bar t$ and $\bar b$)
quarks.    As the angle between the $t$ and $b$ quarks increases, the
velocity of the $b$ quark decreases, the $b$ quark dead cone becomes wider,
the magnitude of the radiation decreases, and the radiation pattern becomes
more isotropic.

\section{SUMMARY}

In this paper we have calculated the soft gluon distribution in $t \bar t$
production, taking the decay of the top quarks fully into account. Although
our primary interest is in hadronic $t \bar t$ production, we have also
considered  several other processes with different colour structures. This
allows us to quantify the  contributions from the different antennae which
determine the radiation patterns.

We have seen how,  in general, the radiation pattern depends on the colour
flow in the $2\to 2$ process, the orientation of the final state particles
with respect to the beam and to each other, and on the top decay width. The
latter controls the interference between radiation before and after the top
decay and between radiation in two decays. When the decay width is large,
the $b$ and $\bar b$ appear almost  instantaneously and they can radiate
coherently, as though produced directly.

We have focussed in particular on two types of distributions. The
distribution in polar angle of the gluon with respect to the beam is
sensitive to  radiation off the initial state particles, and also to the
presence of antennae linking the initial and final state particles (which
are absent in $s$-channel colour-singlet exchange processes). When all the
final state particles are in the transverse plane, the distribution in the
gluon azimuthal angle is more sensitive to the radiation off the final
state particles, in particular to the dead cones of the $t$ and $b$ quarks.

The study of soft gluon radiation in $t \bar t$ events is important not
only for theoretical reasons. The effects we have calculated  are
responsible for the distribution of soft hadrons and jets accompanying the
final state $b$'s and $W$'s. These in turn could play an important role in
the  determination of   the top mass from the reconstructed invariant mass
of its decay products.  For example, the interference between emission {\it
before} and {\it after}  top decay constitutes  an irreducible uncertainty
in such a procedure.  It is important that Monte Carlos used to correct for
these effects in the data contain all the relevant underlying physics.

%
\acknowledgements

This work was supported in part by the UK Science and Engineering  Research
Council. We are grateful to Yuri~Dokshitzer and Lynne~Orr
for useful discussions.

\newpage
%
%
%
%

%
\newpage
%
%
\begin{figure}
\caption{Dependence of the non-factorizing functions $X$ (solid lines) and
$Y$ (dashed lines) for $gg\to{t \bar t}$ on  the centre-of-mass scattering
angle of the $t$ quark, for beam energies $E_{\rm beam} = m_t$ (threshold),
$2m_t$, and $3m_t$.}
\label{FIG:XYFUN}
\end{figure}
%
\begin{figure}
\caption{Colour flow diagrams for the processes
(a) $e^+e^- \to t \bar t$,
(b) $q \bar q \to t \bar t$, and
(c),(d) $gg \to t \bar t$.}
\label{FIG:COLOR}
\end{figure}
%
\begin{figure}
\caption{(a) The orientation of the three-momentum  vectors for the
$dN/d\cos\theta_g$ distributions.  All three-momentum vectors are chosen to
lie in the $xz$-plane.  The $\hat y$-axis is perpendicular to the page. (b)
The orientation of the three-momentum vectors for the $dN/d\phi_g$
distributions.  The three-momentum vectors of the outgoing partons are
chosen to lie in the $xy$-plane.  The three-momentum vectors of the
incoming partons are along the  $\hat z$-axis which is perpendicular to the
page.}
\label{FIG:XYZPLANE}
\end{figure}
%
\begin{figure}
\caption{Distribution of the soft gluon radiation in the polar angle
$\theta_g$; $dN/d\cos \theta_g$ is plotted versus $\theta_g$ for a soft
gluon with energy $\omega_g = 5$~GeV.  Curves are shown for the five
processes listed in  Eq.~(3). The energy of the incoming partons is $E_{\rm
beam} = 2m_t = 280$~GeV, where $m_t = 140$~GeV is the top quark mass.
The Standard Model Born value has been used for the top quark width.  The
directions of the incoming and outgoing partons are indicated at the top of
the figure.  The energies and angles are measured in the parton
centre-of-mass frame.}
\label{FIG:A}
\end{figure}
%
\begin{figure}
\caption{Same as Fig.~\protect{\ref{FIG:A}} but now the top
quark width has been set to
$\Gamma_t = \infty$, which is equivalent to $b\bar b$ production.}
\label{FIG:C}
\end{figure}
%
\begin{figure}
\caption{$dN/d\cos \theta_g$ versus $\theta_g$ for the process $gg \to t\bar
t$ for four values of the top quark width: $\Gamma_t = \infty, 5$~GeV,
0.67~GeV (Standard Model Born value), and 0.  The energy of the soft gluon
is $\omega_g = 5$~GeV and the energy of the incoming gluons  is $E_{\rm
beam} = 2m_t = 280$~GeV. The directions of the incoming and outgoing
partons are indicated at the top of the figure.   The energies and angles
are measured in the parton centre-of-mass frame.}
\label{FIG:E}
\end{figure}
%
\begin{figure}
\caption{$dN/d\cos \theta_g$ versus $\theta_g$ plotted in polar coordinates.
The radiation pattern for the process $gg \to t\bar t$ is illustrated for
$\Gamma_t = \Gamma_{\rm sm}^{(0)}$ (solid line) and $\Gamma_t = \infty$
(dashed line).  The energy of the soft gluon is $\omega_g = 5$~GeV and  the
energy of the incoming gluons is $E_{\rm beam} = 2m_t = 280$~GeV. The
directions of the incoming and outgoing partons are indicated at the top of
the figure.  The energies and angles are measured in the parton
centre-of-mass frame.}
\label{FIG:G}
\end{figure}
%
\begin{figure}
\caption{Distribution of the soft gluon radiation in the azimuthal  angle
$\phi_g$; $dN/d\cos \phi_g$ is plotted versus $\phi_g$ for a soft gluon
with energy $\omega_g = 5$~GeV.  Curves are shown for the five processes
listed in  Eq.~(3). The energy of the incoming partons is $E_{\rm beam} =
2m_t = 280$~GeV, where $m_t = 140$~GeV is the top quark mass.  The Standard
Model Born value has been used for the top quark width.  The directions of the
incoming and outgoing partons are indicated at the top of the figure.  The
$b \> (\bar b)$ quark is $90^\circ$ away from the $t \> (\bar t)$ quark.
The energies and angles are measured in the parton centre-of-mass frame.}
\label{FIG:H}
\end{figure}
%
\begin{figure}
\caption{Same as Fig.~\protect{\ref{FIG:H}} but
now the $b \> (\bar b)$ quark is
$45^\circ$ away from the $t \> (\bar t)$ quark.}
\label{FIG:I}
\end{figure}
%
\begin{figure}
\caption{Same as Fig.~\protect{\ref{FIG:H}}
but now the top quark width has been set
to $\Gamma_t = \infty$, which is equivalent to $b\bar b$ production.}
\label{FIG:L}
\end{figure}
%
\begin{figure}
\caption{$dN/d\phi_g$ versus $\phi_g$ for the process $qq \to t\bar t$ for
four values of the top quark width: $\Gamma_t = \infty, 5$~GeV, 0.67~GeV
(Standard Model Born value), and 0.  The energy of the soft gluon is
$\omega_g = 5$~GeV and the energy of the incoming gluons  is $E_{\rm beam}
= 2m_t = 280$~GeV. The directions of the outgoing partons are indicated at
the top of the figure.   The energies and angles are measured in the parton
centre-of-mass frame.}
\label{FIG:N}
\end{figure}
%
\begin{figure}
\caption{$dN/d\phi_g$ versus $\phi_g$ plotted in polar coordinates. The
radiation pattern for the process $gg \to t\bar t$ is illustrated for
$\Gamma_t = \Gamma_{\rm sm}^{(0)}$ (solid line) and $\Gamma_t = \infty$
(dashed line).  The energy of the soft gluon is $\omega_g = 5$~GeV and  the
energy of the incoming gluons is $E_{\rm beam} = 2m_t = 280$~GeV. The angle
between the $t \> (\bar t)$ and $b \> (\bar b)$ quarks is  $0^\circ,
45^\circ, 90^\circ, 135^\circ$, and $180^\circ$ in parts  (a), (b), (c),
(d), and (e), respectively. The directions of the outgoing partons are
indicated by arrows  on the figure.  The energies and angles are measured
in the parton centre-of-mass frame.}
\label{FIG:P}
\end{figure}
%
%
\widetext
\begin{table}
\caption{The $c_i$ coefficients of Eq.~(\protect{\ref{EQ:GENERAL}})
for $t\bar t$ production,
for $SU(N)$ colour with $C_F = (N^2-1)/(2N)$.}
\begin{tabular}{|l|c|c|c|c|c|}
\hline
process & $c_1$ & $c_2, c_5$ & $c_3, c_4$ & $c_6$ & $c_7, c_8$ \\
\hline
$e^-e^+ \to t \bar t$ & 0 & 0 & 0 & $2C_F$ & $-C_F$ \\
$q \bar q \to Z^* \to t \bar t$ & $2C_F$ & 0 & 0 & $2C_F$ & $-C_F$ \\
$gg\to H\to t \bar t$ & $2N$ & 0 & 0 & $2C_F$ & $-C_F$ \\
$q \bar q \to t \bar t$ & $-{1\over N}$ & $2C_F-{1\over N}$ & ${2\over N}$ &
$-{1\over N}$ & $-C_F$ \\
$gg\to t \bar t$ & $-2C_F+2N+2Y$ & $C_F-X-Y$ & $C_F+X-Y$ & 2Y & $-C_F$ \\
\hline
\end{tabular}
\label{TAB:COLOUR}
\end{table}
%
%
\end{document}